# Lazy Updating increases the speed of stochastic simulations

Kurt Ehlert[1,2] and Laurence Loewe[1,2,a]

[1]*Laboratory of Genetics, University of Wisconsin-Madison, Madison, WI 53706, USA*

[2]*Wisconsin Institute for Discovery, University of Wisconsin-Madison, Madison, WI, 53715, USA*

Oct 7, 2013

**Abstract**

Biological reaction networks often contain what might be called 'hub molecules', which are involved in many reactions. For example, ATP is commonly consumed and produced. When reaction networks contain molecules like ATP, they are difficult to efficiently simulate, because every time such a molecule is consumed or produced, the propensities of numerous reactions need to be updated. In order to increase the speed of simulations, we developed 'Lazy Updating', which postpones some propensity updates until some aspect of the state of the system changes by more than a defined threshold. Lazy Updating works with several existing stochastic simulation algorithms, including Gillespie's direct method and the Next Reaction Method. We tested Lazy Updating on two example models, and for the larger model it increased the speed of simulations over eight-fold while maintaining a high level of accuracy. These increases in speed will be larger for models with more widely connected hub molecules. Thus Lazy Updating can contribute towards making models with a limited computing time budget more realistic by including previously neglected hub molecules.

[a] Author to whom correspondence should be addressed. Electronic mail: loewe@wisc.edu

## I. INTRODUCTION

The behavior of a reaction network is difficult to predict if the network is complex: numerous parts interact at many different rates, so human intuition can be easily fooled. A common approach to understand such complexity is to simulate time series that show how the amounts of important parts in a reaction network change over time. Such simulation algorithms can be applied to a broad range of systems, from models of biochemical reaction networks to models in epidemiology and ecology. Since simulations can be used in many contexts, instead of referring to molecules and reactions, we will use the terms "parts" and "actions". Depending on the context, parts may be molecules in a cell, individuals in a population, or organisms in an ecosystem. The parts in a system are connected with each other through actions, each of which consumes and/or produces specific parts and occurs at a defined rate. Again, depending on the context, actions may be molecular reactions in a cell, births in a population, or predator prey interactions.

Time series that help us to understand these systems can be computed by translating the respective mathematical models into a specific sequence of instructions that can be performed by a computer. If we can assume that all parts in the system are always present in very large numbers, then we can use ordinary differential equations (ODEs) for producing the time series. A number of well established algorithms exist for such ODE models[1]. A major advantage of ODE simulations is that they are very fast. Treating certain models as deterministic makes sense, specifically when the amounts of parts are *very* large, as effectively infinite amounts reduce the stochastic noise to zero. Once the amount of each part can no longer be treated as effectively infinite, stochastic effects become important[2,3]. This is especially dramatic for very small population numbers, because the indivisible nature of molecules in biochemistry or individuals in ecology is not properly handled by ODEs. It has long been known that particular stochastic simulation algorithms (SSAs) provide statistically exact time series for mathematical models that conform to the assumptions behind Continuous Time Markov Chains (CTMCs)[4,5]. While these algorithms are exact, they are also slow compared to ODEs, especially when simulating large complex systems. This problem is compounded by the stochastic nature of the simulation results, because a statistical understanding of the stochastic process requires many simulations to accurately estimate values such as the mean and the variance. Speeding up stochastic simulations without losing numerical accuracy has been an active area of research for some time[2,6-16].

To use an SSA, we need to provide it with a model describing all the parts and their initial amounts, as well as all the actions and their dynamics. With a fully specified model and a stream of random numbers, an SSA can simulate time series of the model's behavior. These time series are almost always unique for complex models, because (pseudo)random numbers are used to decide (i) which action happens next, and (ii) when it occurs. This makes it very unlikely that sequences of actions and their time are identical for two independent runs of a complex simulation model. Each choice in such stochastic simulations is governed by a well-defined probability distribution describing a waiting time for when the corresponding action is expected to occur next. These waiting times are always exponentially distributed for CTMC models. The exponential distribution takes a single parameter, which we call the 'propensity'. We will present the definition of the propensity in the *Methods* section (equation 1), as it can help understand Lazy Updating.

A common feature of SSAs is that they exploit the mathematics of exponential distributions in clever ways to quickly determine which action will happen next, while staying true to the CTMC nature of the model. This action is not necessarily the one with the shortest waiting time; rather, all SSAs compute the propensity for each action to be proportional to the probability that this particular action will occur next. To calculate the propensity of an action at a given point in time, we *only* need to know (i) the equation that describes its dynamics, and (ii) the amounts of those parts in the system that affect this propensity. For our study it is important to note that the propensity of an action only needs to be updated when there was a change in the amount of any part that is needed for that reaction to occur.

If a part is consumed and produced by many actions, then we call the part 'highly connected'. Highly connected parts are common in biology. For example, there are 'hub' proteins[17] and molecules such as ATP and water that take part in many reactions. If a model contains highly connected parts, then propensity updates require much more computing time, since many actions change the amount of these hub parts and many action propensities need to be updated after every single occurrence of any one of these actions. This makes it very expensive to include common molecules like water, ATP, NADP into simulation models. For researchers with a fixed budget of CPU-time, this can mean that certain aspects of biological models will remain unexplored, such as aspects of energy metabolism that require monitoring ATP, an important energy currency in cells.

Here we present the "Lazy Updating" mechanism, an extension for SSAs, which was developed to reduce the computational cost associated with including hub parts. Lazy Updating allows researchers to incorporate highly connected parts into models without sacrificing large amounts of computational speed. In brief, Lazy Updating skips propensity updates until the amounts of the parts change by more than a specified percentage. We tested Lazy Updating on two example models. We chose a linear pathway with one highly connected molecule to represent a model that we expected to show a substantial increase in computational efficiency. To explore the limits of the precision of Lazy Updating, we chose a simple model of exponential population growth, where we expected Lazy Updating to perform poorly. Based on our results, Lazy Updating substantially speeds up simulations of models with hubs, while also maintaining a high level of accuracy for parts that follow less than ideal dynamics.

## II. METHODS

### A. Lazy Updating algorithm

An action's propensity typically depends on the amount of reactant parts; therefore, to keep track of an action's exact propensity, the propensity should be updated every time the amount of a reactant changes. If an SSA updates the propensities *every* time any reactant amount changes, we say that the SSA uses 'Immediate Updating'. Instead of updating propensities every time the reactant amounts change, Lazy Updating postpones a propensity update until the amount of a reactant has accumulated changes up to some percentage of the old value. Users of Lazy Updating specify the tolerated differences as a relative error percentage, and a lower percentage means that propensity updates occur more often.

Lazy Updating fits well into the general structure of SSAs and hence can work with a broad range of SSAs including Gillespie's Direct Method[5], the Next Reaction Method[2,3,11], and others. An SSA with Lazy Updating works almost the same way as an SSA without Lazy Updating as can be seen in the outline of our implementation of Lazy Updating (see pseudocode in Figure 1). The major difference between the two is in the step that updates propensities (see bold lines in Figure 1).

We implemented Lazy Updating with the Sorting Direct Method[4,5,14], which is based on Gillespie's Direct Method. Implementing Lazy Update can be a very simple modification as seen in Figure 1, if each part already has a list of the actions it affects, because lazily updated parts must know which propensities depend on their amount. To avoid updating propensities more than once per simulation step, we use a stack $S$ and a vector of bits $B$ as follows. $S$ stores pointers to actions whose propensities need updating, while $B$ stores for each action, whether it is already scheduled for updating. When an action occurs, *S* is passed to all of the affected parts, and the parts tell *S* which propensities need to be updated. Before a part does that, it queries *B* to determine, which propensities are already scheduled for an update in order to avoid propensities from being updated more than once per simulation step. Once all dependent propensities are added to the stack, the SSA goes through the stack and updates all of the propensities indicated by the stack's elements, and then resets the vector of bits.

This strategy is similar to tau-leaping strategies that have been used to speed up simulations. Instead of holding each species count fixed between jumps as in classical tau-teaping[12], we just hold a subset of them fixed until a pre-defined accuracy criterion demands an update.

**B. Test models**

To test the performance and accuracy of Lazy Updating, we chose the two artificial models shown in Figure 2. We deliberately chose simple models to help focus attention on the core principles of Lazy Updating. It is easy to see that the benefits of Lazy Updating will be model dependent and increase for more complex models with larger numbers of highly connected hub parts. Propensities of actions with mass action dynamics are calculated as

$$\lambda_k(x) = r_k \prod_i \left( \frac{x_i!}{(x_i - v_{k,i})!} \right) \tag{1}$$

where $\lambda_k$ is the propensity of action *k*, $r_k$ is the mass action rate constant of action *k*, $x_i$ is the amount of part *i*, and $v_{k,i}$ is the stoichiometric change of part *i* in action *k*.

**Model A** was designed to test the potential for speeding up simulations by Lazy Updating and is shown in Figure 1A. It represents a linear unidirectional pathway of 102 reactions, with a steady source at the beginning, a sink at the end and a series of many identical mass action reactions in between. The specialty of this model is that every other reaction consumes or produces ATP, making this molecule a highly connected hub like in more complicated biological models. In our model, the initial amount of ATP is 5000, and the initial amount of each intermediate molecule named $X_i$ is 1000. The rate constants for the first and last actions are $r_1 = r_{102} = 100$, and the rate constants of all other actions are $r_k = 0.0003$.

**Model B** was designed to highlight potential problems with the accuracy of simulation results produced by Lazy Updating. We expect these problems to be biggest when molecules are always only consumed or produced; in such circumstances the errors from Lazy Updating never cancel out. Thus we expect the simple exponential growth model shown in Figure 1B to highlight important limits for the accuracy of results generated by Lazy Updating. The rate of the single action is 0.4. The initial molecule amount is 10.

### C. Simulations

Simulations were all performed by the SSA algorithm know as the 'Sorting Direct Method' (SDM) described by McCollum et al[2,6-16]. We implemented the SDM in C++ measured the simulation CPU run time by calculating the number of clock cycles used by the simulation with the C++ standard library *clock()* function. We also recorded the number of propensity updates skipped by Lazy Updating.

To ensure the statistical independence of our simulations we used the Mersenne Twister[17,18] pseudorandom number generator (PRNG) with the exponential and uniform random variate generators available from http://www.boost.org (boost version 1.52.0). The PRNG was seeded by the C++ standard function *time* and generated all pseudorandom numbers without reseeding during an a series of simulations. *time* returns the number of seconds since 00:00 hours on January 1$^{st}$, 1970. We only reseeded for different series of simulations, not for each individual simulation.

**D. Accuracy measurements**

Lazy Updating requires the user to specify a tolerance, which is used to determine how often propensity updates occur and which impacts the accuracy of time series that report the amounts of parts. To test the accuracy of these reported amounts, we set the Lazy Update tolerance to 1%, an unnecessarily high value chosen to exaggerate potential accuracy problems that a poorly chosen Lazy Updating tolerance may cause. According to our speed test results such a 1% tolerance is much higher than necessary to achieve most of the increase in computational speed that can result from Lazy Updating (see next section). To make this test for accuracy even harsher, we wanted to see the size of errors on time series accuracy that this poorly chosen large tolerance would cause in Model B, a model that we expect to be particularly sensitive to such errors, as reactions *always* move away from the last value.

To compare the accuracy of 5,000 Lazy Updating simulations with 5,000 precise simulations (Immediate Updating), we recorded the amounts of the parts every 0.1 time units over a period of 200 or 20 time units for Models A and B, respectively, and calculated the mean and standard deviation of the amounts at every recorded time point.

To further examine how closely time series generated with Lazy Updating match those generated by Immediate Updating, we calculated the percent difference, *d*, between the amounts observed in 5,000 Lazy Updating and 5,000 Immediate Updating time series every 0.1 time units. For each observed time point,

$$d = \frac{100 \, |\mu_I(t) - \mu_L(t)|}{\mu_I(t)} \qquad (2)$$

where $d$ is the percent difference, $\mu_I(t)$ is the mean amount observed by Immediate Updating at time $t$, and $\mu_L(t)$ the mean observed by Lazy Updating at the same time. To reduce the noise in the results, we reported $d_{10}$, which we define as the rolling mean of $d$, averaged over a window spanning 100 time units on each side and thus including 200 observations.

To assess whether 5000 simulations had enough statistical power to reliably detect a difference between Lazy Updating and Immediate Updating in Model B, we conducted 1000 parametric and non-parametric statistical tests that compared two groups of surrogate results, each with a sample size of $n = 5000$ appropriately drawn random numbers. These numbers were drawn from two normal distributions $N(\mu, \sigma)$, parameterized with a mean ($\mu$) and standard deviation ($\sigma$) as observed at the end of the simulations for Model B when run by Lazy Updating ($\mu_L = 28047.77$, $\sigma_L = 8677.42$) and Immediate Updating ($\mu_I = 28648.63$, $\sigma_I = 9095.23$). We used a two-sided $t$-test as parametric test (implemented by *t.test* in R) and a two-sided Mann-Whitney test as non-parametric test (implemented by *wilcox.test* in R). We found that Lazy Updating with the unnecessarily sloppy tolerance of 1% led to differences that could be expected to be statistically significant when comparing 5000 simulation results, but not when comparing 1000 simulation results (see Table 1). These values are obviously model dependent, so high precision simulations of stochastic systems that wish to employ Lazy Updating will need to analyze the accuracy they require before running final simulations. To provide a point of comparison, we added to Table 1 corresponding estimates for Model A using $\mu_L = 7885.62$, $\sigma_L = 49.28$, $\mu_I = 7884.13$ and $\sigma_I = 50.07$.

**E. Estimating speedup from Lazy Updating**

To measure the speedup caused by Lazy Updating, we recorded the speed of simulations with Immediate Updating and with a range of Lazy Updating tolerances. Then we calculated the mean and standard deviation of the CPU time required for each simulation, as well as the average number of skipped propensity updates per propensity update. We found that $S$, the speed increase relative to Immediate Updating is well approximated by

$$S = \frac{T_I}{T_L} = \frac{T_I}{T_R + T_P/(k+1)} \tag{3}$$

where, $T_I$ is the time it takes to run a simulation with Immediate Updating, $T_L$ is the time it takes to run a simulation with Lazy Updating, which is the sum of $T_R$, the average time needed for everything not related to propensity computation (such as updating part amounts, drawing random numbers, etc.) and the time spent on updating propensities, which can be computed from $T_P$, the average time spent on updating propensities when zero propensity updates are skipped and $k$, the average number of skipped updates between actually performed propensity updates.

Obviously, the parameters in equation (3) will be model specific and here we only considered our Model A. For a given Lazy Update tolerance, we can measure $T_I$, $T_L$ and hence $S$, just by observing the simulations. To measure $k$ we added a counter to the code, which provided an average of how many updates were skipped. To estimate the other parameters, we measured the average of the above parameters based 10 Immediate Update runs and 10 runs with a very relaxed Lazy Update setting of 0.2%, which we assume approximates the maximal speedup possible for propensity computation (i.e. $T_P / (k+1)$ goes to 0). In that case we can approximate $T_R$ by our observation of $T_L$ and use the resulting equation (3) to approximate $S$ for less tolerant Lazy Update settings.

## III. RESULTS

### A. Speed increase for Model A: pathway with hubs

As shown in Figure 3, Lazy Updating substantially increased the speed of simulations of Model A representing a linear pathway with ATP as hub molecule (Figure 2A). Simulations that used a tolerance of 0.2% ran over eight times faster than simulations that used Immediate Updating, which is equivalent to a tolerance of 0%. We found that equation (3) approximates the expected speed increase with high accuracy, if we used high quality estimates for the corresponding model-specific parameters (Method 1); using quick and easy estimates of the same parameters (Method 2, see above) turns equation (3) in a rough but still useful rule of thumb. We plotted both estimates against the actual speed increase as observed in our simulations (Figure 3).

### B. Accuracy for Model A: the near equilibrium case

To assess the accuracy with which Lazy Updating computes time series of amounts of parts, we compared time series generated by Lazy Updating and Immediate Updating. We used an unnecessarily high Lazy Update tolerance of 1% to test the robustness of Lazy Update in light of less than perfect tolerance choices. Since all time series are realizations of a stochastic process, we compared them by calculating means and standard deviations every 0.1 time units from 5,000 Lazy Updating and 5,000 Immediate Updating time series. The mean of the Lazy Updating time series very closely overlaps with the mean of the Immediate Updating time series, and the standard deviations of the time series also match so closely that differences are almost invisible when plotted on top of each other (Figure 4; the non-overlapping red and blue areas are barely visible at a zoom of >1000%). This precision is achieved despite our choice of a 1% Lazy Update tolerance that is about 10 times more sloppy than would be recommended based on Figure 3.

Figure 5 further quantifies a local average difference in amounts between Lazy Update and Immediate Update simulations. We usually found this difference to be near 0.01% in our simulations. While this difference could in principle be caused by our use of Lazy Update, it could also stem from random stochastic effects in these independent sets of stochastic simulations. To quantify the impact of such stochastic effects, we calculated the percent difference between the set of 5,000 Immediate Updating time series used above and an independent set of 5,000 equivalent Immediate Updating time series (Figure 5). The two equivalent sets of Immediate Updating time series had about the same percent difference as when we compared the Lazy Updating and Immediate Updating time series. We interpret this as an indication that most of the percent difference is due to stochastic effects. Unless many more simulations are run, we will not be able to effectively distinguish between completely precise results and results generated by Lazy Updating for our Model A (see also the analysis of statistical power for Model A in Table 1). This precision increases our confidence that Lazy Updating closely approximates Immediate Updating for a large range of biologically interesting simulation scenarios that involve near equilibrium conditions for hub parts.

### C. Accuracy for Model B: non-equilibrium growth

To test the accuracy limits of the Lazy Update method, we next looked for a more challenging model. Specifically, we looked for a model where different neglected updates would not cancel each other out and hence generate much larger systematic biases over time. We chose Model B, a simple exponentially growing population with only births and no deaths (Figure 2B). The constant growth in this model makes Lazy Updating consistently underestimate propensity, thus consistently delaying actions (and hence further grow) for a little bit each time. These delays can add up as shown here, if the absolute time between the crossing of two thresholds of part amounts is critical.

Starting with an initial population size of 10, the time series of Model B produced by Lazy Updating closely approximated the time series from Immediate Updating at the beginning. Absolute differences increased as the population grew (Figure 6). However, relative errors remained small (Figure 7). Even at the end of the simulation with mean population sizes of 28047.77 and 28648.63, the absolute difference was only 600.9 (~2.1%) between the means of distributions of final amounts as observed at the end of time series generated with Lazy Updating versus Immediate Updating. These differences between the two distributions are statistically significant as determined by the two-sided parametric *t*-test or the more robust non-parametric two-sided Mann-Whitney-*U*-test if 5000 simulations are compared, but not significant if 'only' 1000 simulations are compared (see Table 1).

We conducted a statistical power analysis to investigate, how many simulations would be necessary to detect such differences with any statistical accuracy (Table 1). We found that differences of this size are expected to be significant if 5000 simulations are compared (as we did above), but not necessarily if only 1000 simulations are compared (see Methods Section). These analyses are highly model dependent as can be seen when comparing Models A and B in Table 1. Thus, it will be up to modelers using Lazy Update to conduct statistical power analyses if very precise results need to be observed.

A few causes conspire to produce the differences caused by Lazy Update that we reported for Model B above: (i) as for Model A, here we chose an unnecessarily excessive Lazy Update tolerance of 1%, which is at least 10 fold higher than more reasonable values like 0.05% or 0.1%; these deliver almost all the desired computing speedup with much smaller impacts on accuracy; (ii) Lazy Update errors do not cancel out under growth in Model B as they do under the near equilibrium conditions in Model A where amounts are decreased and increased by different simultaneously acting processes; (iii) these errors would not be visible if we had computed substantially less stochastic repeats; in such cases they are likely to be irrelevant as they will be drowned in stochastic noise unless studies require very high precision and compute very large numbers of repeats to that end.

The causal analysis of processes that affect the choice of Lazy Update tolerance levels and corresponding accuracy implications is sufficient here as more detailed analyses necessarily depend on details of models and are thus best conducted in that context.

**IV. DISCUSSION**

Proper analyses of models using stochastic simulation algorithms have broad applications, but regularly stretch computing time allocations. This generates a need for algorithmic improvements to speed up computations without sacrificing significant accuracy. We presented the Lazy Updating extension for SSA algorithms (Figure 1) and demonstrated an over eight-fold increase in simulation speed (Figure 3) for a medium sized model with a 'hub' molecule like ATP that is produced and consumed in many reactions (Figure 2A). This speed increase required a Lazy Update tolerance of 0.1% and did not substantially decrease the accuracy of the simulations (Figures 4, 5, Table 1). Based on our results, a Lazy Updating tolerances between 0.1% and 0.05% are good values, but the best values are model-dependent. The speedup is achieved by avoiding many costly propensity updates that do not contribute much to overall simulation accuracy. Speed increases are expected to be larger for bigger and more connected models.

**Applicability.** The accuracy results of Lazy Updating for the pathway model (Figure 4, 5) suggest that Lazy Updating is particularly well suited for "hub" parts that are often consumed and produced. Lazy Updating can speed such simulations up significantly without a noticeable reduction of accuracy when compared to Immediate Updating as we demonstrated in our model. These improvements are well positioned to help enable simulations of many small molecules that exist in high numbers in cells and are involved in many diverse reactions. Such molecules like ATP are essential for a cell's energy budget, implying that genes affecting levels of these molecules might be under relatively strong selection. Investigating such questions is facilitated by simulation methods that are both quick and precise.

**Accuracy.** To test Lazy Updating's accuracy under harsh conditions, we used it with a high tolerance on the birth-only growth Model B, which we expected to lead to a particularly large loss of accuracy. Since the size of the population only increases, Lazy Updating consistently underestimates related propensities, and the absolute amount of underestimation gets worse over time. We used a high tolerance of 1%, which was much more than necessary to obtain a good speedup from Lazy Updating. Despite the large tolerance, Lazy Updating only generated minor errors (Figure 6) that would not have been detected if it were not for the large numbers of stochastic repeats we computed. Equation 3 and Figure 3 show that most of the speed increase is obtained by very small tolerances, which means that we can obtain most of the speed increase without sacrificing much accuracy.

Errors are larger, if Lazy Updating is applied to parts with large dynamical changes, where either production or consumption of these parts dominates and errors from Lazy Updating no longer cancel out. In such cases, local errors in amounts are only limited by the specified Lazy Update tolerance. The same is not true of global errors as seen in Model B: if production is autocatalytic, then all resulting slight changes in propensity combine to affect future amounts. It is easy to see that similar conclusions also apply to autocatalytic consumption and many other situations, where dynamic changes lead to substantial and sustained uni-directional fluxes of parts.

While the combination of many small local errors can lead to larger errors in absolute amounts of affected parts, these larger errors only matter if reaching certain precise threshold amounts will trigger major events, which have an extremely precise timing that is critical for some important system property. In such special models minute differences can easily blow up into major effects. To deal with this appropriately requires a detailed analysis of the corresponding dynamics and is thus outside of the scope of this present study. This problem is not unique to Lazy Updating, as other approximate stochastic simulation algorithms, such as tau leaping[5,12], would also struggle to precisely simulate the timing of a major event that is caused by very small differences in amounts of some part that crossed a threshold. ODE solvers have dedicated 'root-finding' features for accurately detecting thresholds[19].

**Tolerance.** Lazy Updating requires a user-specified parameter for the tolerance. While we do not have a principled method for choosing a 'correct' tolerance we found that even small tolerance values, such as 0.1% or 0.05%, can lead to large speedups while maintaining a high level of accuracy. We suggest testing small values for the tolerance, running some example simulations to test the speedup and accuracy, and then choose the smallest value that gets most of the speedup. The choice of tolerance can be accelerated by using equation (3) and estimating its parameters to the desired precision. As described in the *Methods* Section, this might be done as little as two simulations; one simulations would use Immediate Updating and the other would use a large Lazy Updating tolerance (e.g. 1%). This allows us to assume that approximately no time is spent on propensity updates, specifying all parameters in equation (3) and making it ready for prediction.

Since the tolerance is currently specified as a relative percentage, the number of skipped updates scales with the current number of molecules. When the number of molecules is low, and accurate propensities become especially important, propensity updates occur more often. When the number of molecules is very high, and accurate propensities depend more on other rare molecules and not on the abundant hub molecule, then propensity updates occur less often. This behavior of Lazy Updating is likely to be advantageous, because it dynamically readjusts the accuracy and speedup based on the current state of the system. Other types of tolerance can be introduced easily if needed (e.g. absolute tolerances) and since tolerances are accessed on a part-by-part basis, it is possible to control the use of Lazy Updating as finely as conceivably useful.

**Conclusion.** Lazy Updating trades a small amount of accuracy for a substantial speedup. Furthermore, implementing Lazy Updating within existing stochastic simulation algorithms is relatively simple. We expect that Lazy Updating will be extremely useful for situations where a model is large, highly connected, and requires many simulations. For example, the important problems of parameter estimation and model selection frequently require a large number of simulations[20]. If parameter estimation or model selection involves a large, highly connected reaction network, then Lazy Updating may drastically alleviate the computational burden and enable the exploration of new biological questions that involve network hubs in biochemistry and ecology.

**Acknowledgements:**

We thank David F. Anderson and James R. Faeder for stimulating discussions and helpful comments on this problem. This work was supported by NSF CAREER Award 1149123 to LL and by the Wisconsin Institute of Discovery at the University of Wisconsin-Madison.

Table 1: *P*-value quantiles for 1000 tests of the null-hypothesis that the amounts obtained by Lazy Updating and Immediate Updating at the end of our simulations are drawn from the same distribution, even though they are actually drawn from different distributions as parameterized by simulation results for Model B ($\mu_L = 28047.77$, $\sigma_L = 8677.42$, $\mu_I = 28648.63$, $\sigma_I = 9095.23$) and Model A ($\mu_L = 7885.62$, $\sigma_L = 49.28$, $\mu_I = 7884.13$ and $\sigma_I = 50.07$). Results significant at the 5% level are bold. Dots indicate the same entry as above.

| Model | Test | *n* | *P*-value 10% quantile | *P*-value median | *P*-value 90% quantile |
|---|---|---|---|---|---|
| B | t-test | 5000 | **3.34 x 10$^{-6}$** | **6.91 x 10$^{-4}$** | **0.03072** |
| . | . | 1000 | 0.0593 | 0.1295 | 0.6908 |
| . | Mann-Whitney | 5000 | **4.11 x 10$^{-6}$** | **9.7 x 10$^{-4}$** | **0.03835** |
| . | . | 1000 | **6.96 x 10$^{-3}$** | 0.1475 | 0.7433 |
| A | t-test | 5000 | **6.04 x 10$^{-3}$** | 0.1202 | 0.6849 |
| . | . | 1000 | 0.0522 | 0.3984 | 0.8749 |
| . | Mann-Whitney | 5000 | **5.50 x 10$^{-3}$** | 0.1361 | 0.7613 |
| . | . | 1000 | 0.0510 | 0.3897 | 0.8540 |

```
0  // Pseudocode of SSA with Lazy Updating
1  set initial conditions for the SSA model: amounts, propensities, times;
2  initialize S, an empty stack for storing pointers to actions;
3  initialize B, a vector with one bit for each existing action ID integer;
4  while ( Time_Current < Time_End ) {
5     choose next action A and Time_To_Next_Action;
6     Time_Current += Time_To_Next_Action ;                // advance time
7       for each part P whose amount is changed by A {
8          Will_Update_Propensities = true ;
9          // next check if update can be lazy:
10         if ( ( P allows Lazy Update) and
11              ( | Amount_Change | < Amount_At_LastPropensityUpdate * ε ) )
12         then
13            Will_Update_Propensities = false ;
14         endif
15         Amount_Current += Amount_Change ;      // change state of system
16         // next schedule updates only once
17         if ( Will_Update_Propensities ) {
18           for each action X in all actions affected by P {
19             if( B [ X.getID() ] == 0) then
20                 B [ X.getID() ] =  1 ;
21                 add pointer to action X to stack S ;
22             endif
23           } // goto 18
24         endif
25      } goto 7
26      update propensities of all actions not allowed to be lazy;
26      update propensities of all actions in S;
27      clear S; reset all entries in B to zero;
28 } // goto 4
```

FIG 1. Pseudocode of the Lazy Updating algorithm. The bold lines denote the Lazy Update additions embedded in an underlying standard SSA algorithm (non-bold). The main idea is to tolerate small inaccuracies in propensities caused by minute changes in the amount of parts as detected in lines 8-14. These inaccuracies are tolerated until they reach a threshold, at which all relevant propensities are recomputed (line 26). To make the algorithm efficient, the code should store a propensity dependency graph for each part that shows which actions have propensities dependent on the part's amount. In addition, more complicated checks (Line 16-24) can avoid potential multiple updates of the same propensity, which can further improve speed, but is not necessary. Variable names start with capital letters and getID() returns the unique ID of an action.

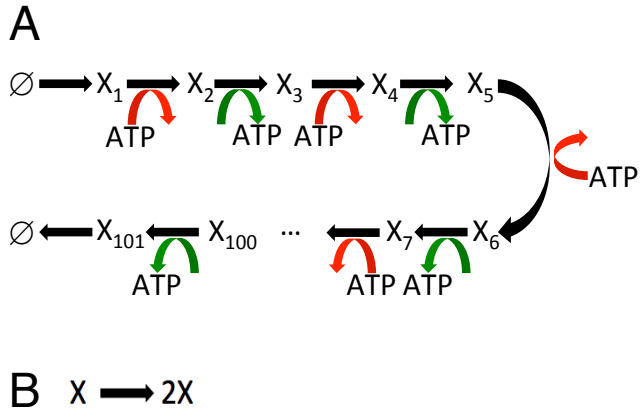

FIG. 2. Models used to test the Lazy Updating algorithm.

(A) Model A: The ATP model used to test the speed of the Lazy Updating algorithm. Green arrows indicate production, red arrows consumption. ATP in this model is a well-connected hub and is subject to Lazy Updating; thus, when the amount of ATP changes, we update affected propensities according to the Lazy Update algorithm. If the amounts of the $X_i$ molecules change, then we update any affected reaction propensities immediately.

(B) Model B: The birth process model used to test the accuracy of Lazy Updating for parts with amounts that are continuously growing. Since the molecule count only increases, Lazy Updating consistently underestimates the reaction propensity. Consequently, the accuracy of Lazy Updating for this system should be substantially worse than for most other systems, where Lazy Updating will be closer to the true value more often as increasing and decreasing actions can cancel out.

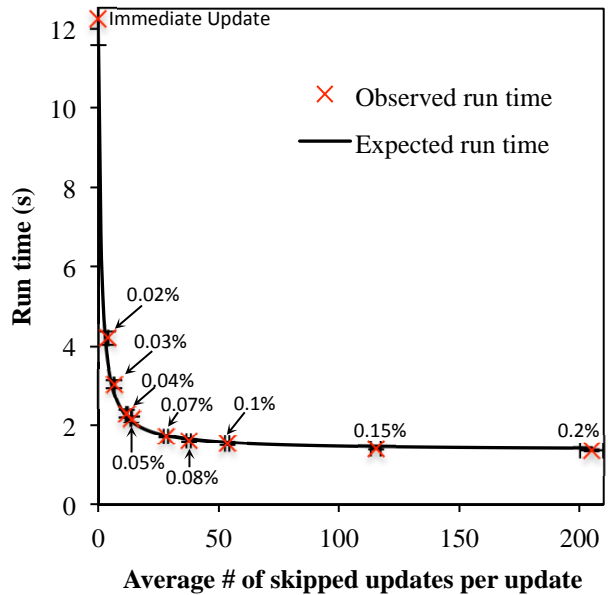

FIG. 3. A comparison of the observed and expected speed increases for the linear pathway Model A (Figure 2A), where only the hub part is subject to Lazy Updating. Each red cross denotes a measured point of comparison averaging over 10 independent simulations, where black error bars indicate standard deviations of the run time and the number of skipped propensity updates, respectively. Note that error bars are barely visible, even for sample sizes of only 10 simulations. The lines are calculated by equation (3) in the main text and denote the expected increase in speed from Lazy Updating. For our model, the parameters were estimated to be $T_I = 12.25$, $T_R = 1.37$, $T_P = 10.88$ (see Methods Section for details).

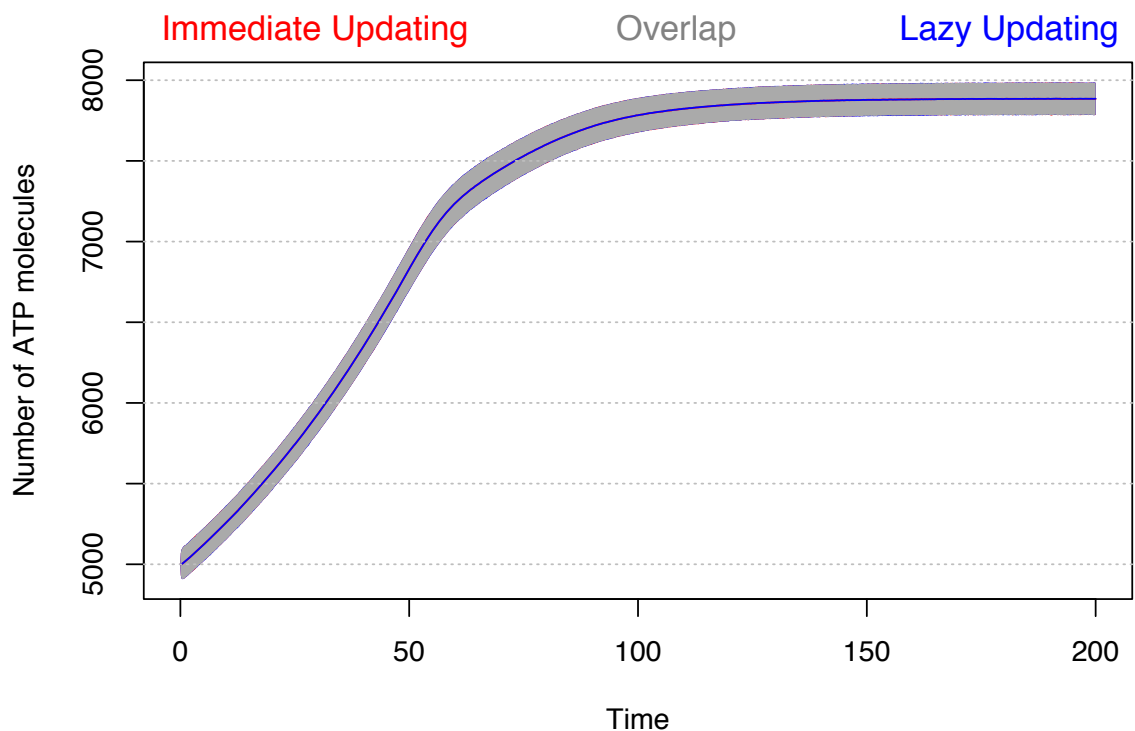

FIG. 4. Comparison of time series results obtained by Lazy Update to full precision SSA for the linear pathway Model A (Figure 2A). The 5000 Lazy Update simulations in this plot ran more than eight-fold faster than the corresponding 5000 fully precise Immediate Update simulations. Shown are aggregated time series as described in the Methods Section. Blue denotes approximate results from Lazy Updating, Red denotes precise results from Immediate Updating. The grey area is where the blue and red areas overlap. Note that these curves are so similar that the red area and curve hide completely behind the blue in this plot. Lines show the mean of simulated time series. Areas indicate two standard deviations around the mean. The 1% Lazy Update tolerance used here was more imprecise than would be recommended for production use based on Figure 3.

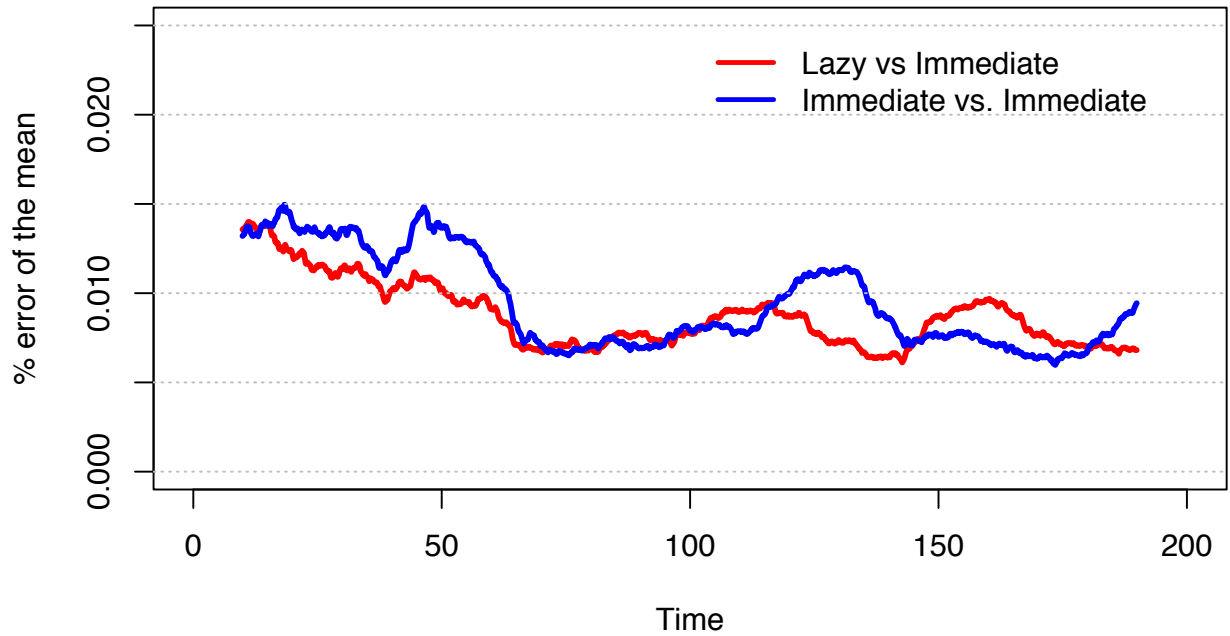

FIG. 5. Relative accuracy of Lazy Update results for linear pathway Model A. The red line is the rolling mean of the percent difference between the mean of 5,000 Lazy Update simulations and the mean of 5,000 Immediate Update simulations. The blue line is the rolling mean of the percent difference between the mean of 5,000 Immediate Update simulations and the mean of another set of independent 5,000 Immediate Update simulations. The rolling means used windows of 10 time units, purely to reduce the visual clutter in the plot. The percent difference is calculated by equation 2 in the main text. This figure is based on the data in Figure 4, supplemented by an additional set of Immediate Update simulations.

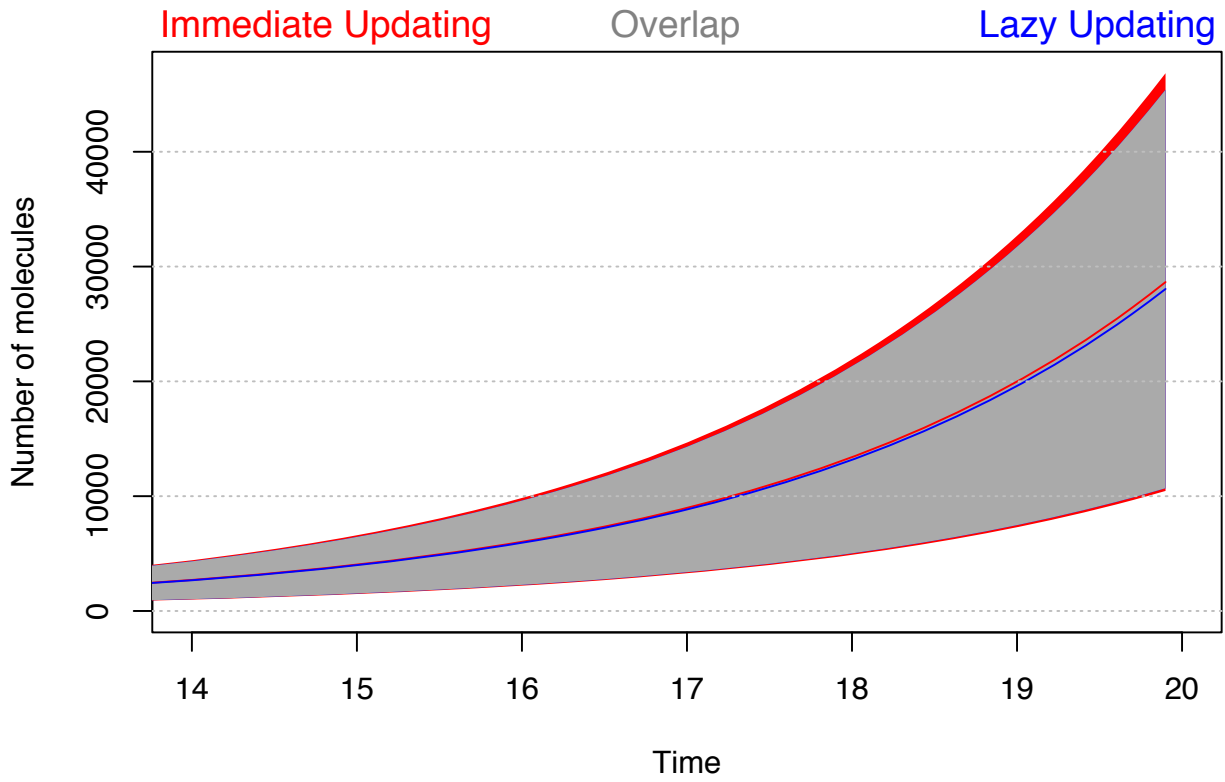

FIG. 6. Accuracy test of Lazy Update simulations for the birth process Model B (Figure 2B), which is expected to suffer particularly badly from Lazy Updating. For details please see Figure 4, which is identical except that it was run for a different model. We compared the 5000 measurements of numbers of parts at the end of the simulation by Lazy Updating with those measured by Immediately updating. Using a *t*-test, we find $P = 2.18 \times 10^{-4}$, whereas a Mann-Whitney-U Test returned $P = 8.26 \times 10^{-4}$; thus, the differences reported at the end of runs for Model B are significant, if compared in 5000 stochastic runs, but not in 1000 stochastic runs.

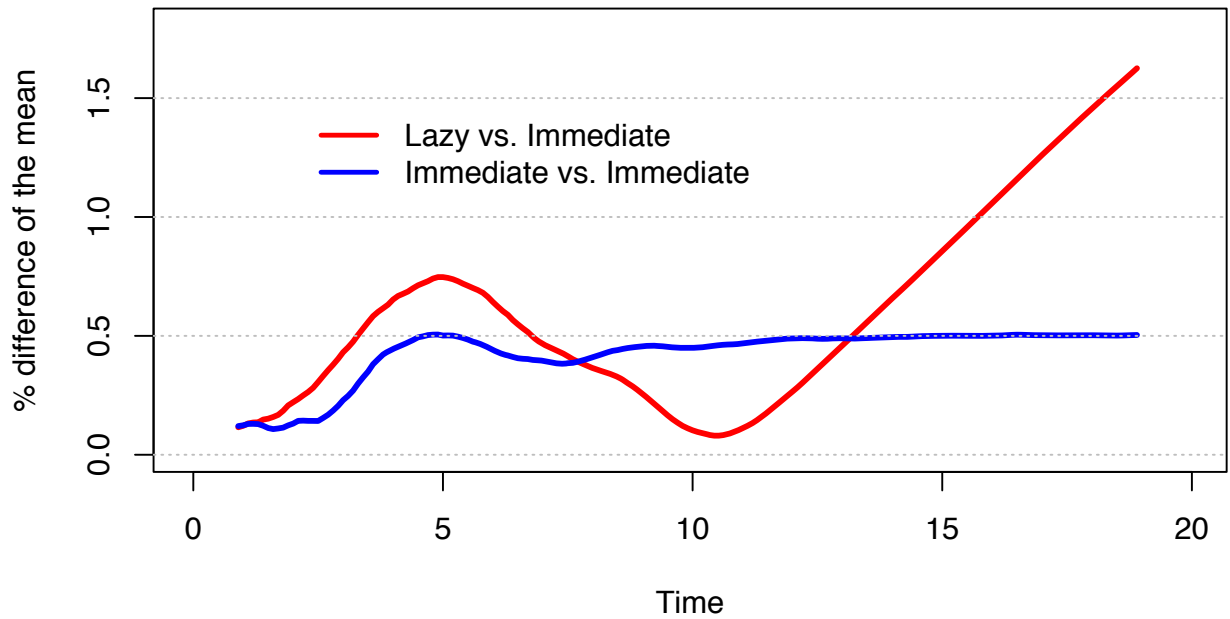

FIG. 7. Accuracy test of Lazy Update simulations for the birth process Model B (Figure 2B), which is expected to suffer particularly badly from Lazy Updating. For details on this plot please see Figure 5, which is identical except that it was run for a different model.